\documentclass[prb,showpacs,preprintnumbers,preprint]{revtex4}
\usepackage{graphicx}
\begin{document}
\def\rhov{{\mbox{\boldmath{$\rho$}}}}
\def\tauv{{\mbox{\boldmath{$\tau$}}}}
\def\Lambdav{{\mbox{\boldmath{$\Lambda$}}}}
\def\sigmav{{\mbox{\boldmath{$\sigma$}}}}
\def\xiv{{\mbox{\boldmath{$\xi$}}}}
\def\oh{{\scriptsize 1 \over \scriptsize 2}}
\def\of{{\scriptsize 1 \over \scriptsize 4}}
\def\tf{{\scriptsize 3 \over \scriptsize 4}}
\title{Ferroelectricity Induced by Incommensurate Magnetism}

\author{A. Brooks Harris}

\affiliation{ Department of Physics and Astronomy, University
of Pennsylvania, Philadelphia, PA 19104}

\date{\today}

\begin{abstract}
Ferroelectricity has been found to occur in several insulating systems,
such as TbMnO$_3$ (TMO) and Ni$_3$V$_2$O$_8$ (NVO) which have more
than one phase with incommensurately modulated long-range magnetic order.
Here we give a phenomenological model which relates the symmetries of
the magnetic structure as obtained from neutron diffraction to the
development and orientation of a spontaneous ferroelectric moment
induced by the magnetic ordering.  This model leads directly
to the formulation of a microscopic spin-phonon interaction which
gives explains the observed phenomena.  The results are given in
terms of gradients of the exchange tensor with respect to
generalized displacements for the specific example of NVO.
It is assumed that these gradients will now be the target of
first-principles calculations using the LDA or related schemes.
\end{abstract}
\maketitle

\section{Introduction}

\begin{figure} [h]
{\vspace {0.1 in} \includegraphics[width=6.5cm]{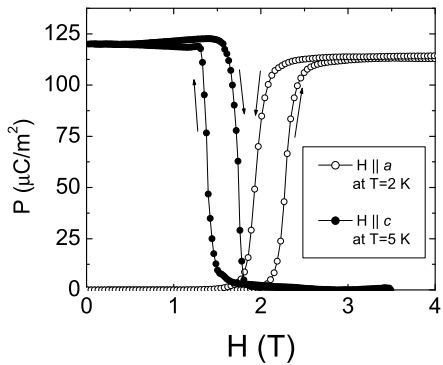}}
\hspace {1.5 in}
\caption{\label{GAVIN} Adapted from Ref. \protect{\onlinecite{FERRO}}.
Spontaneous polarization of NVO versus magnetic field applied along
the $a$ and $c$ axes.}
\end{figure}

Recently studies have focussed on a
a family of multiferroics which display a phase transition
in which there simultaneously develops long-range incommensurate
magnetic and uniform ferroelectric order. The most detailed
studies have been carried out on the system Ni$_3$V$_2$O$_8$
(NVO).\cite{Rogado,LawesKenzelmann,FERRO,EXPT}  A similar analysis of
TbMnO$_3$ (TMO) has also appeared.\cite{TMO1,TMO2}
(For a review of both systems see, see Ref. \onlinecite{REV}.)
To illustrate the phenomenon we show in Fig. \ref{GAVIN} data for the
spontaneous polarization as a function of applied {\it magnetic}
field.

\noindent
This data indicates a strong coupling between the magnetic
order parameters and the polarization order parameters.
To interpret the effect of this coupling, we reproduce in Fig.

\begin{figure} [ht]
\includegraphics[width=6.0 cm]{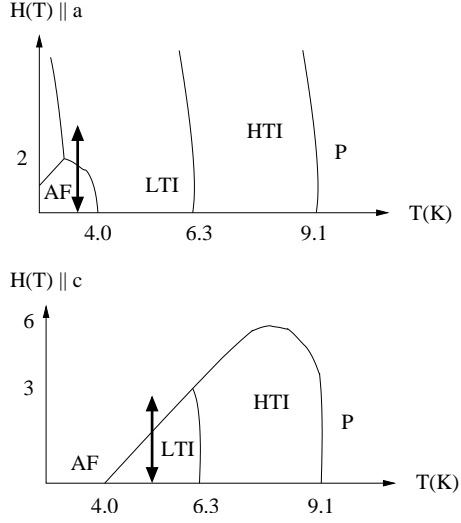}
\caption{\label{PD} Adapted from Ref. \protect{\onlinecite{LawesKenzelmann}}.
Phase diagram of NVO for magnetic fields up to 8 T
applied along the {\bf a} direction (top) and along
the {\bf c} direction (bottom). P (AF) denotes the paramagnetic
(antiferromagnetic) phase.  For $H || {\bf c}$ there is no phase
boundary between the P and AF phases. The arrowed
vertical lines indicate
the paths followed in Fig. \protect{\ref{GAVIN}}.}
\end{figure}

\noindent
\ref{PD} the phase diagram as function of temperature $T$ and magnetic field
$H$ applied along the crystal {\bf a} and {\bf c}-axes.  Here we show
the three magnetic phases that occur for $T \geq 2$K. The HTI
and LTI phases are two distinct incommensurate phases.  A spontaneous
polarization appears throughout the LTI phase, but does not
appear in the other phases.  Thus, applying a strong enough $H$
to cross the LTI-AF phase boundary will kill the spontaneous
polarization for ${\bf H} || {\bf c}$ or allow it for ${\bf H} || {\bf a}$.
Since this transition is discontinuous, the hysteresis
shown in Fig. \ref{GAVIN} is to be expected.

This phenomenon has been explained\cite{FERRO} on
the basis of a phenomenological model which invokes a Landau
expansion in terms of the order parameters describing the
incommensurate magnetic order and the order parameter describing
the uniform spontaneous polarization.  Already from this
treatment it was clear that a microscopic model would have to
involve a trilinear interaction Hamiltonian proportional to the
product of two spin variables and one displacement variable.
Furthermore, the symmetry requirements of the phenomenological model
would naturally be realized by a proper microscopic model.
In this paper we summarize the symmetry analysis and briefly
describe the microscopic formulation which underlies the
symmetry analysis.

\section{Symmetry of the Magnetic Phases}

We first discuss the crystal structure of NVO.  The space group
of NVO is Cmca (\#62 in

\begin{figure} [ht]
{\vspace {0.1 in} \includegraphics[height=4.5cm]{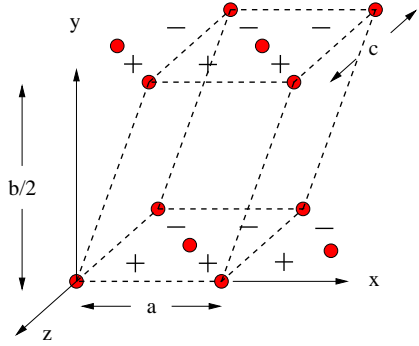}}
\hspace {0.05 in}
\caption{\label{UC} Positions in NVO of Ni cross-tie sites (filled circles)
and Ni spine sites (indicated by $+$ and $-$ signs according to whether
their $y$ coordinate is larger than or less than that of the
layer). The dashed lines outline the primitive unit cell.}
\end{figure}

\noindent
Ref. \onlinecite{HAHN}). Apart from
the primitive translations ${\bf a}_1 = (a/2) \hat i + (b/2) \hat j$,
${\bf a}_2 = (a/2) \hat i - (b/2) \hat j$, and ${\bf a}_3=c \hat k$),
the generators of the space group may be taken to be $2_x$
a two-fold rotation about the $x$-axis, $m_x$, a reflection taking
$x$ into $-x$, and $m_z$ a glide plane which takes $z$ into $-z$,
followed by a translation of $(b/2)\hat j + (c/2) \hat k$.  The
primitive unit cell contains two formula units of NVO 
and the magnetism is due to the Ni ions (see Fig. \ref{UC}).
There are two crystallographically inequivalent Ni sites which
we will refer to as "spine" and "cross-tie" sites.  
The positions of the spine sites within the unit cell are
${\bf r}_1 = (a/4,d,c/4)$, ${\bf r}_2=(a/4,-d,3c/4)$,
${\bf r}_3= (3a/4,-d,3c/4)$, and ${\bf r}_4=(3a/4,d,c/4)$.
The parameter $d\approx 0.13b$ indicates that the a-c planes are buckled
kagom\'{e}-like planes.  The cross-tie sites within the unit cell are 
at ${\bf r}_5=(0,0,0)$ and ${\bf r}_6=(a/2,0,c/2)$.

\begin{figure} [ht]
\includegraphics[width=6.0 cm]{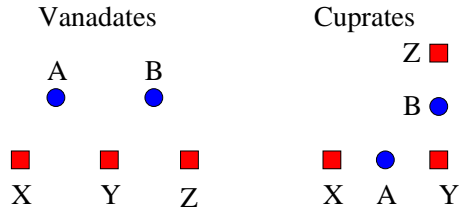}
\caption{\label{JJ} First and second neighbor exchanges paths 
(respectively X-A-Y or Y-B-Z and X-A-B-Z) in
the vanadates\protect{\cite{Rogado}}
(left) and cuprates (right), where A and B are oxygen ions
and X, Y, and Z are magnetic ions.  In the vandates the
first neighbor exchange path is nearly a 90$^{\rm o}$ path, so that
the nn exchange is weak, whereas in the cuprates the path is a
$180^{\rm }$ path, so that here $J_2/J_1 \sim 0.1$ (see Ref.
\protect{\onlinecite{2342}}), whereas for the vanadates
$J_2/J_1$ is much larger.\protect{\cite{EXPT}}}
\end{figure}

The Ni spins form weakly coupled chains (or spines) parallel to the
${\bf a}$-axis.  The nearest neighbor (nn) and next-nearest neighbor
(nnn) antiferromagnetic interactions ($J_1$ and $J_2$, respectively)
along the spine compete with one another because, as shown in Fig.
\ref{JJ}, in contrast to the situation in the cuprates, in the vanadates
the nn interaction is anomalously weak.  A result of this
competition is that the spins order in an incommensurate
state whose wavevector, $q$, along the ${\bf a}$ axis is given
by\cite{Nagamiya}
\begin{eqnarray}
\cos (aq/2) = - J_1/(4J_2) \ .
\end{eqnarray}
As the temperature is lowered (at zero magnetic field), the
system first orders in the HTI phase.  Here the spins develop order
mainly along the easy ({\bf a}) axis so that their magnitude
varies sinusoidally. At a lower temperature the
LTI phase is entered in which there is additional ordering
of the transverse spin components to more nearly satisfy
the constraint of fixed spin length.

\begin{figure} [h]
\vspace{0.1 in} \includegraphics[height=5.5cm]{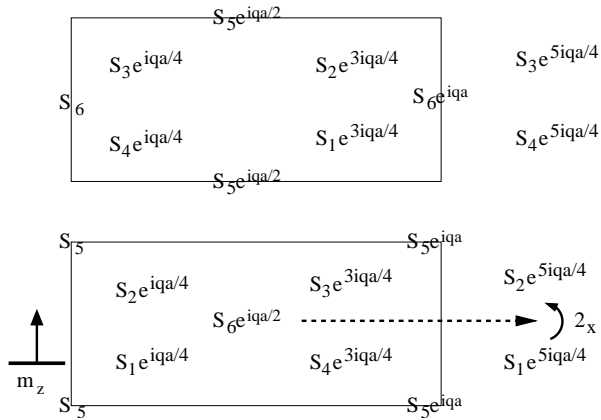}
\caption{\label{SPSTR} Two adjacent $x-z$ planes in NVO (buckling not shown).
Complex spin amplitudes for wavevector $q$ (whose real part gives the
spin moment).  Here $2_x$ is the axis of two-fold rotation about $x$ and
$m_z$ represents the glide plane which consists of the mirror plane
shown at $z=c/4$ together with the displacement $b/2$ along $y$
(from one plane to the next).}
\end{figure}

The above description of the incommensurate phases is too
simplistic.  Accordingly we now turn to a more detailed
discussion which takes proper account of the structure within
each unit cell.
The $\alpha$-component of the spin order at position ${\bf R}$
(which is the $\tau$th site in the unit cell) is expressed in terms
of Fourier components at wavevector $q$ as
\begin{eqnarray}
S_{\alpha \tau} ({\bf R}) &=& [ S_{\alpha \tau} ({\bf q})
e^{i {\bf q} \cdot {\bf R}} + S_{\alpha \tau} ({\bf q})^*
e^{-i {\bf q} \cdot {\bf R}}]/2 \ ,
\end{eqnarray}
where $S_{\alpha \tau} ({\bf q})$ is a complex-valued Fourier
coefficient and $S_{\alpha \tau} (- {\bf q})=S_{\alpha \tau}({\bf q})^*$.
In Fig. \ref{SPSTR} we show the complex quantity $S_{\alpha \tau}({\bf q})
e^{i {\bf q} \cdot {\bf R}}$ whose real part gives the spin order at
position ${\bf R}$. Assuming only a single Fourier wavevector
condenses (see below), we see that the spin structure is characterized
by the six vector Fourier components of the unit cell.  Since these
quantities are three component vectors and each components is
complex-valued, this structure is characterized by
36 real-valued parameters.  As we shall see, the use of symmetry
drastically reduces the number of parameters needed to characeterize
the magnetic structure.

Translational invariance indicates that the
Landau expansion of the free energy at quadratic order
in terms of these Fourier components must assume the form
\begin{eqnarray}
F_2 &=& \sum_{\bf q} F_2({\bf q}) \ ,
\end{eqnarray}
where
\begin{eqnarray}
F_2({\bf q}) &=& \sum_{\alpha \alpha'; \tau \tau' }
c_{\alpha \tau; \alpha' \tau' ; {\bf q} }
S_{\alpha \tau} (-{\bf q}) S_{\alpha' \tau'} ({\bf q} ) \ .
\end{eqnarray}
The eigenvalues of the quadratic form $F_2({\bf q})$
are the various inverse susceptibilities.  We are
interested in the stability of the quadratic form (since we assume
the ordering transition is a continuous one).  We may plot the
least positive (least stable) eigenvalue as a function of wavevector
$q$.

\vspace{0.6 in}
\begin{figure} [h]
{\vspace {0.1 in} \includegraphics[height=4cm]{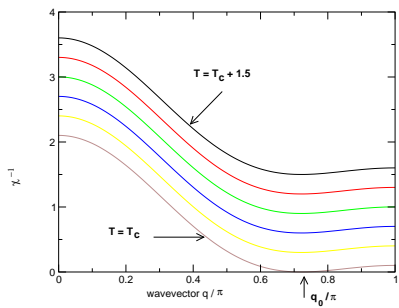}}
\hspace {1.5 in}
\caption{\label{chi} Inverse susceptibility as a function of wavevector
for a sequence of temperatures for the $J_1$, $J_2$ model with
$J_1/J_2=2.56$ (as for NVO). Temperature is in units of $J_1$.}
\end{figure}
\noindent
Such plots for a sequence of temperatures are shown in Fig. \ref{chi},
where one sees that some value of wavevector is selected at which the
paramagnetic phase first becomes unstable relative to the
formation of long-range order.  Of course, this value of $q$
is determined by the interactions between spins, which are
not well known.  Here we are interested
in what can be said
on the basis of symmetry.  A cardinal principle is that we
always reject the possibility of accidental degeneracy.  That is,
there is only a single value of $q$ which becomes unstable as
the temperature is lowered from the paramagnetic phase. (This phenomenon 
is called {\it wavevector selection}.)  For a ferromagnet,
the selected wavevector is $q=0$ and for a simple cubic antiferromagnetic
$q=(\pi/a)(1,1,1)$, where $a$ is the lattice constant. For
NVO the situation is more complicated because along each
spine one has competing nn and nnn
antiferromagnetic interactions which cause the inverse susceptibility
to have its minimum at some nonspecial value of wavevector.  This
value is\cite{LawesKenzelmann} ${\bf q} \approx 0.28 (2 \pi /a) \hat i$.
The unit cell contain several spins, so that the actual incommensurate
spin configuration can realize different symmetries within the unit cell.
In the HTI phase one has incommensurate spin ordering
mainly on the spines in which the spins are oriented along the
easy (${\bf a}$) axis with sinusoidally varying amplitude.  In
the LTI phase, in addition to the order preexisting in the HTI phase,
significant spin order transverse to the ${\bf a}$
axis appears, mainly on the cross-tie sites.  Superficially,
this type of structure may seem difficult to characterize.  However,
by focussing on the symmetry of the structure we are able to
obtain a convincing phenomenological analysis of the
symmetry of the magnetoelectric interaction.

We therefore now turn to the an analysis of the magnetic symmetry.  Since
the wavevector lies along the $x$-axis, the free energy, $F_2({\bf q})$
must be invariant under the operations of the crystal which
leave the $x$-axis invariant.  The group of these symmetry operations\
includes the identity operation ${\cal E}$, a two fold rotation
about the $x$-axis, $2_x$, the glide reflection through a $z$-plane,
$m_z$ and $2_xm_z$.  (Some of these are illustrated in Fig. \ref{SPSTR}.)
Without going through the details of group theory, one can say (in this
case) that the eigenvector of the quadratic form $F_2({\bf q})$, must also
be an eigenvector of the operation $2_x$ (with eigenvalue either $+1$
or $-1$) and also of $m_z$ (with eigenvalue either $+1$ and $-1$).
Thus the structure determination is much simplified.  Instead of
having to determine the 6 complex vectors ${\bf S}_n$ of Fig. \ref{SPSTR},
we use the fact that these spin components form a vector which is
simultaneously an eigenvector of $F_2({\bf q})$, $2_x$, and $m_z$.
Accordingly we may specify (or guess) the
four possible sets of eigenvalues of $2_x$ and $m_z$.  This set
of eigenvalues is technically known as the {\it irreducible
representation} $\Gamma$, or irrep, for short.  The actual active irrep
is selected as the one which best fits the diffraction data.
For the HTI phase the active irrep is found to be\cite{LawesKenzelmann}
$\Gamma_4$ for which the
eigenvalues of $2_x$ and $m_z$ are $-$ and $+$, respectively.
For this irrep we have to specify the
3 complex spin amplitudes of a single spine site and the $y$ and $z$
components of a single cross-tie site.  The other amplitudes are
then determined by application of the symmetry operators
$2_x$ and $m_z$.  Since we reject accidental degeneracy, the ordering
of the HTI phase can only involve a single irrep.  In Fig.

\begin{figure} [h]
\vspace {0.1 in} \includegraphics[width=8.5cm]{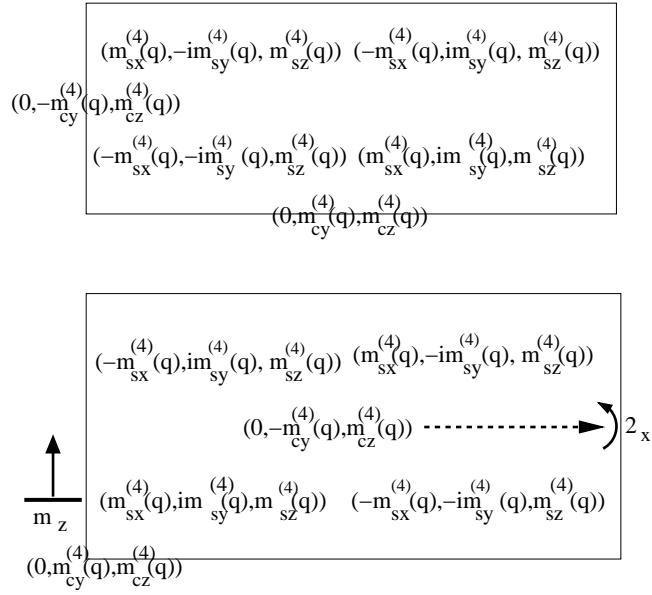}
\caption{\label{G4} As Fig. \protect{\ref{SPSTR}}.
Fourier coefficients $S_{\alpha \tau}(q)$ in terms of
symmetry-adapted parameters for the HTI irrep $\Gamma=\Gamma_4$,
for which the eigenvalues of
$2_x$ and $m_z$ are $-1$ and $+1$, respectively.}
\end{figure}

\noindent
\ref{G4}, we show the spin function for the HTI phase of NVO
corresponding to the eigenvalues of $2_x$ and $m_z$ respectively
$-1$ and $+1$.\cite{LawesKenzelmann}  Notice that the characterization
of the magnetic structure is reduced to the specification of
only five complex-valued {\it symmetry adapted coordinates}.

\begin{figure} [h]
\vspace{0.2 in} \includegraphics[width=8.5 cm]{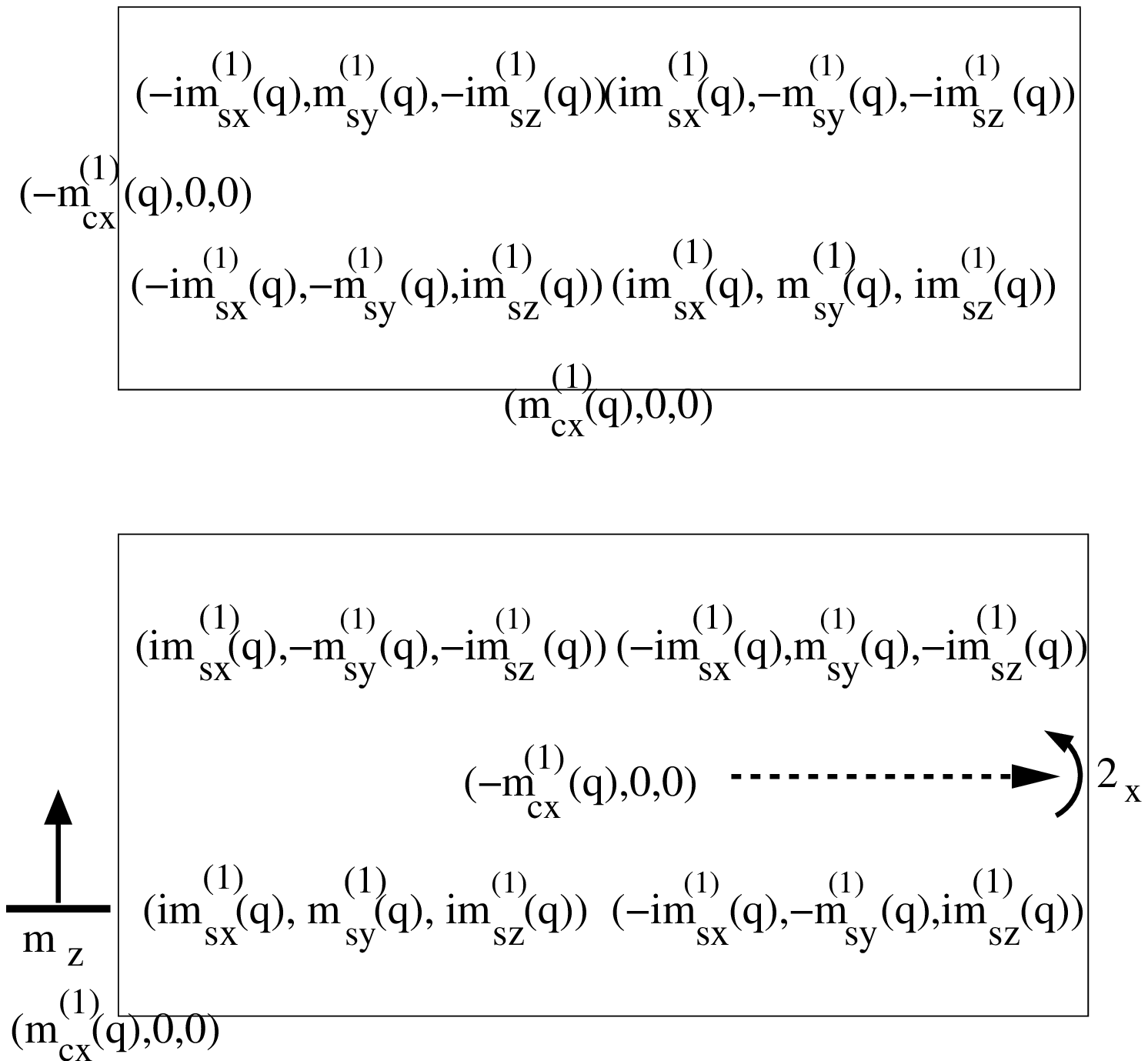}
\caption{\label{G1} As Fig. \protect{\ref{SPSTR}}.
Fourier coefficients $S_{\alpha \tau}(q)$ in terms of
symmetry-adapted parameters for the LTI irrep $\Gamma=\Gamma_1$,
for which the eigenvalues of
$2_x$ and $m_z$ are both $+1$.}
\end{figure}

When the LTI phase is entered, the situation is much the same.
In addition to a spin function with the HTI symmetry, we now 
introduce additional ordering which has a different symmetry.
This additional symmetry is found to correspond to the eigenvalues
of $2_x$ and $m_z$ respectively $+1$ and $+1$ (of irrep $\Gamma_1$)
and the corresponding spin functions (which require the specification of
four additional symmetry adapted coordinates) are shown in Fig. \ref{G1}.

The cystal NVO is inversion symmetric and we have not yet discussed
how this affects the analysis.  Under spatial inversion ${\cal I}$
the components of spin transform into each other.  Since spin is
a pseudovector, inversion does not change the orientation of the spin,
but takes it into the spatially inverted position and therefore
takes $q$ into $-q$ (which is the same as complex conjugation).
Also inversion takes the site ${\bf r}_1$ in the unit cell into
site ${\bf r}_3$ and site ${\bf r}_2$ into site ${\bf r}_4$.
Referring to Figs. \ref{G4} and \ref{G1}, one see that we have defined the
spin components $m_{s\alpha}^{(n)}$ with appropriate phase factors of
$i$ so that\cite{FOOT}
\begin{eqnarray}
{\cal I} m_{s\alpha}^{(n)} &=& {m_{s\alpha}^{(n)}}^* \ . 
\label{IMEQ} \end{eqnarray}
This same relation also holds for the cross-tie order parameters
without the necessity of introducing phase factors.  Now
consider the expansion of the free energy in terms of these variables,
where for notational convenience we set (for each irreducible
representation) $\xi_1(q)=m_{sx}(q)$,
$\xi_2(q)=m_{sy}(q)$, $\xi_3(q)=m_{sy}(q)$, and $\xi_4(q)$, and
if need be, $\xi_5(q)$ are similar cross-tie order parameters. 
In terms of these variables the quadratic free energy is
\begin{eqnarray}
F_2({\bf q}) &=& \sum_{n,n';\Gamma} c_{n,n'}^\Gamma \xi_n^\Gamma(q)^*
\xi_{n'}^\Gamma (q) \ ,
\end{eqnarray}
where the superscript identifies the irreducible representation.
To ensure the reality of $F_2({\bf q})$ we must have 
$c_{n',n}^\Gamma={c_{n,n'}^\Gamma}^*$.  Equation (\ref{IMEQ}) implies that
\begin{eqnarray}
{\cal I} \xi_n^\Gamma (q) = \xi_n^\Gamma (q)^* \ .
\label{IEQ} \end{eqnarray}
Now for $F_2({\bf q})$ to be invariant under ${\cal I}$ we must have
\begin{eqnarray}
{\cal I} {\cal } F_2({\bf q}) = \sum_{n,n';\Gamma} c_{n,n'}^\Gamma
[{\cal I} \xi_n^\Gamma (q)^*]
[{\cal I} \xi_{n'}^\Gamma (q)] = \sum_{n,n'} c_{n,n'}^\Gamma 
\xi_n^\Gamma (q) \xi_{n'}^\Gamma (q)^* = F_2({\bf q}) \ .
\end{eqnarray}
This indicates that $c_{n,n'}^\Gamma=c_{n'n}^\Gamma$, which together with
$c_{n',n}^\Gamma (q)=c_{n,n'}^\Gamma (q)^*$
means that all the coefficients $c_{n,n'}^\Gamma$ are
{\it real} valued.  This means that apart from an overall phase
factor, all the components of the eigenvector are {\it real}.
So for each representation, $\Gamma$, we write
\begin{eqnarray}
\xi_n &=& \sigma e^{i \phi} r_n \ ,
\label{REAL} \end{eqnarray}
where the $r_n$ are the components of a real valued unit vector.\cite{FOOT}
Instead of giving the representation $\Gamma$ as a superscript
on the $\xi$'s, $\phi$'s, and $r$'s, we will write HTI (for
$\Gamma_4$) or LTI (for $\Gamma_1$).
By writing the symmetry adapted coordinates in the form of Eq. (\ref{REAL}),
one sees that this type of incommensurate ordering is characterized by an
$x$-$y$ like order parameter $\sigmav (q) \equiv \sigma \exp (i \phi)$
which has a magnitude $\sigma$ and a phase $\phi$.  The magnitude of
the order parameters $\sigma$ are fixed by terms (we have not considered)
which are fourth order in the spin components.   These and higher order terms
have no effect on the normalized eigenvector as long as one is close to the
ordering transition.  From Eq. (\ref{IEQ}) we see that for each representation
\begin{eqnarray}
{\cal I} \sigma = \sigma \ , \ \ \ 
{\cal I} \phi = - \phi \ .
\label{SIGEQ} \end{eqnarray}
Apart from the overall
phase, the spin structure of the HTI phase is specified by the
wavevector q and the 5 real-valued parameters $r_n^{\rm HTI}$
and the additional ordering appearing in the LTI phase requires specifying
4 real-valued parameters $r_n^{\rm LTI}$ and the relative phase
$\phi_{\rm LTI} - \phi_{\rm HTI}$.

\section{Magnetoelectric Coupling: Phenomenology}

In Ref. \onlinecite{FERRO} it was proposed that one could understand
the magnetically induced ferroelectricity in terms of the
following Landau expansion
\begin{eqnarray}
F &=& F_M + {1 \over 2 \chi_E} {\bf P}^2 + V_{\rm int} \ ,
\end{eqnarray}
where $F_M$ is the free energy of the system in the absence of
a nonzero polarization (and which is given up to quadratic order
by $F_2({\bf q})$), ${\bf P}$ is the uniform
electric polarization, and $\chi_E$ is the
electric susceptibility, which we assume to be finite, since we assume
that when the magnetism is absent the spontaneous polarization is zero.
Here $V_{\rm int}$ is the magnetoelectric interaction which is
posited to be of the form
\begin{eqnarray}
V_{\rm int} = \sum_{\alpha \beta \gamma} c_{\alpha \tau; \beta \tau'}(
{\bf q}) S_{\alpha \tau}(-{\bf q}) S_{\beta \tau'}({\bf q}) P_\gamma \ ,
\label{TRIL} \end{eqnarray}
where $\gamma$ labels the Cartesian components of ${\bf P}$.
This interaction may be viewed as that of an effective electric
field proportional to a quadratic combination of spin variables
and which thereby induces a nonzero polarization only when the
magnetic order parameters are nonzero. 
Because the spin components are expressible in terms of the
$x$-$y$-like order parameters introduced in Eq. (\ref{REAL}),
we may write this as
\begin{eqnarray}
V_{\rm int} = \sum_{A B \gamma} c_{A B\gamma}({\bf q})
\sigmav_{\rm A} (q) \sigmav_{\rm B} (-q) P_\gamma \ ,
\end{eqnarray}
where A and B are summed over the values HTI and LTI.  In the HTI
phase the LTI order parameter is zero, so that
\begin{eqnarray}
V_{\rm int} = \sum_\gamma c_{HTI,HTI,\gamma}(q)
|\sigma_{HTI} (q)|^2 P_\gamma \ .
\end{eqnarray}
But since this interaction has to be inversion invariant, the
coefficient $c_{HIT,HTI,\gamma}(q)$ must vanish.  Thus symmetry
does not allow a spontaneous polarization to be induced by
magnetic ordering in the HTI phase. A simple (but not rigorous)
way to reach this conclusion is to observe that the phase of an
incommensurate wave will vanish arbitrarily close to some lattice
site.  If one then takes this lattice site as the origin, the
magnetic structure will have inversion symmetry relative to this
new origin and hence a spontaneous polarization is not allowed. 

In the LTI phase the situation is different.  Although the
term involving $|\sigma_{\rm LTI}|^2$ vanishes by the above argument,
one has
\begin{eqnarray}
V_{\rm int} &=& \sum_\gamma P_\gamma [
c_{\gamma}(q) \sigmav_{\rm HTI}(q) \sigmav_{\rm LTI} (q)^*
+ c_{\gamma}(q)^* \sigmav_{\rm HTI}(q)^* \sigmav_{\rm LTI} (q)] \ .
\end{eqnarray}
Using Eq. (\ref{SIGEQ}) one sees that for this to be invariant under
spatial inversion the coefficient $c_{\gamma}(q)$ has to be pure imaginary,
so that $V_{\rm int}$ has to be of the form
\begin{eqnarray}
V_{\rm int} &=& \sum_\gamma r_\gamma \sigma_{\rm HTI} \sigma_{\rm LTI}
\sin( \phi_{\rm LTI} - \phi_{\rm HTI}) P_\gamma \  ,
\end{eqnarray}
where $r_\gamma$ is real-valued.  Thus a requirement that
for NVO incommensurate magnetic order induce a spontaneous polarization
is that the two order parameters $\sigmav_{\rm HTI}$ and $\sigmav_{\rm LTI}$
should not be in phase, {\it i. e.} $\phi_{\rm HTI} \not= \phi_{\rm LTI}$.
An analysis\cite{ANAL} of the fourth order terms in the Landau expansion of
$F_M$ indicates that indeed these two order parameters do not have the
same phase.  We now consider the effect of the symmetries $2_x$ and
$m_z$.  From the symmetries of the active irreps, (or from inspection
of Figs. \ref{G4} and \ref{G1}) one can see that
\begin{eqnarray}
2_x [\sigmav_{\rm HTI} \sigmav_{\rm LTI}^*] &=&
- [\sigmav_{\rm HTI} \sigmav_{\rm LTI}^*] 
\end{eqnarray}
and
\begin{eqnarray}
m_z [\sigmav_{\rm HTI} \sigmav_{\rm LTI}^*] &=&
[\sigmav_{\rm HTI} \sigmav_{\rm LTI}^*] \ . 
\end{eqnarray}
But the product $P_\gamma \sigmav_{\rm HTI} \sigmav_{\rm LTI}^*$
must be invariant under these two symmetry operations.  Thus
$P_\gamma$ must change sign under $2_x$ but be invariant under
$m_z$.  This implies that ${\bf P}$ can only have a nonzero
component along ${\bf b}$, as is observed.\cite{FERRO}

To summarize: the Landau symmetry analysis for NVO explains why
ferroelectricity can only be induced in the LTI phase which
is described by two different symmetry order parameters.
(This is reminiscent of the ahalysis of the symmetry of 
second harmonic generation.\cite{FEIBIG})  In addition
this symmetry analysis correctly predicts that incommensurate magnetic
order can only induce a spontaneous polarization along
the crystallographic {\bf b} direction.  A completely similar
analysis has been developed\cite{TMO2} for TbMnO$_3$ which exhibits
a similar magnetically induced ferrolectricity.\cite{TMO1}

\section{Magnetoelectric Coupling: Microscopics}

\begin{figure} [h]
{\vspace {0.1 in} \includegraphics[width=7.8cm]{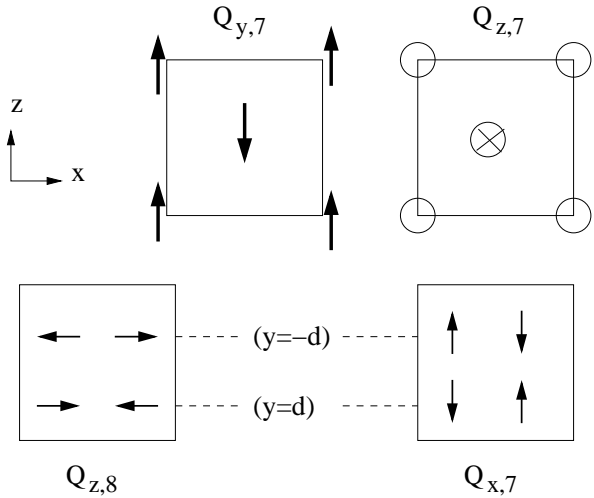}} \hspace {1.5 in}
\caption{\label{PH1} Adapted from Ref. \protect{\onlinecite{MICRO}}.
Top: generalized displacements of cross-tie Ni atoms which transform
like $y$ (left) and like $z$ (right).  Plain circles represent displacements
out of the page and a circle with a cross represents a displacement into
the page.  Bottom: generalized displacements
of Ni spine atoms which transform like $z$ (left) and like $x$ (right).}
\end{figure}

As noted in Ref. \onlinecite{FERRO},
the form of the trilinear magnetoelectric coupling of Eq. (\ref{TRIL})
leads directly to the construction of a microscopic interaction
which must underlie the symmetry analysis presented above.  Since
a polarization can only come from an atomic displacement, the
microscopic interaction must involve the coupling of two
spin operators and one displacement operator.  The generalized exchange
tensor interaction between spins at sites $i$ and $j$ is
\begin{eqnarray}
{\cal H} (i,j) &=& \sum_{\alpha \beta}
J_{\alpha \beta} (i,j) S_\alpha (i) S_\beta (j) \ .
\end{eqnarray}
Here we allow for both symmetric and antisymmetric Dzialoshinskii-Moriya
(DM) interactions.\cite{DM1,DM2} To get a displacement operator, we
simply expand the exchange tensor ${\bf J}$ up to first order in the
atomic displacements.  Note that a uniform polarization must arise
from a zero wavevector optical phonon.  Since the energies and
wavefunctions of the optical phonons are unknown, we will
express the results in terms of the symmetry adapted generalized
displacements (GDs) at zero wavevector.  We only need to consider
GDs which transform like a
vector under the symmetry operations of the crystal, because only
these GDs can carry a dipole moment.  These GDs
are denoted $Q_{\alpha n}$, where $\alpha$ labels the  component
$x$, $y$, or $z$ according to which $Q$ transforms.  Some of these
GDs are easy to construct: a GD in which
all crystallographically equivalent atoms {\it e. g.} Ni spines,
Ni cross-ties, or V ions, move along the same crystallographic
direction $\alpha$ will generate the $Q_{\alpha,n}$'s for
$n \leq 6$, because NVO has six crystallographically inequivalent
sites. The other relevant GDs are given
elsewhere,\cite{REV,MICRO} but we show a few of these in
Fig. \ref{PH1}.  To identify the symmetry of the GDs
shown here, it is useful to consider the effect
of $2_x$, $m_x$, and $2_y$, which is a two-fold rotation about
a $y$-axis passing through a spine site.  For instance, consider the

\noindent
effect of these operations on $Q_{z,8}$.  
This GD is odd under $2_x$ (a two-fold rotation about the center
of the cell), even under $m_x$, and is odd under $2_y$.
These evaluations verify that this GD does have
the same symmetry as a $z$ coordinate.  This
means that, perhaps surprisingly, this mode gives rise to a
dipole moment in the $z$-direction.  One can also see that $Q_{z,8}$
couples to the GD in which all the cross-ties move in
parallel along the $z$-axis. Suppose the atom-atom potential is
repulsive.  Then when the spine atoms move as shown in Fig.
\ref{PH1}, they force the cross-tie at the center of the unit cell
to move towards negative $y$ and positive $z$ and the cross-tie
at the corner of the unit cell to move towards positive $y$
and positive $z$.  Thus $Q_{z,8}$ couples to a uniform displacement
of the cross-tie along the $z$-axis, as its symmetry label
indicates.

In this development the spin-phonon coupling involves the various
derivatives
\begin{eqnarray}
J_{\alpha \beta}^{\gamma_n} & \equiv &
{\partial J_{\alpha \beta} \over \partial Q_{\gamma_n} } \ .
\end{eqnarray}
The algebra needed to construct the spin-phonon Hamiltonian
${\cal H}_{\rm s-}$ when the spin operators are replaced by
their thermal averages is straightforward but tedious and is
given elsewhere when nn spine-spine,
and spine-cross-tie, as well as second neighbor spine-spine
interactions are taken into account.\cite{REV,MICRO}
Only terms which involve $Q_{\alpha,n}$ with $\alpha=y$ are
nonzero, in agreement with the macroscopic symmetry analysis.\cite{FERRO}
To illustrate the results of these calculations, we give here
the result for nn spine-spine interactions\cite{REV,MICRO}
\begin{eqnarray}
{\cal H}_{\rm s-p} &=& 16 N_{uc} \sum_p Q_{y_p} \sum_{\mu , \nu = x,y,z}
\Lambda_{\mu \nu}^{(nn)} \  \Im \left[ m_{s,\mu}^{(4)}(q)^*
m_{s,\nu}^{(1)}(q) \right] \ ,
\label{YPEQ} \end{eqnarray}
where $N_{uc}$ is the number of unit cells, $\Im$ indicates the
imaginary part, and
\begin{eqnarray}
\Lambdav^{(nn)} &=& \left[ \begin{array} {c c c}
J_{xx}^{y_p} c & D_z^{y_p} s & D_y^{y_p}c \\
- D_z^{y_p} s & -J_{yy}^{y_p} c &
- J_{yz}^{y_p} s \\ D_y^{y_p} c &
J_{yz}^{y_p} s& - J_{zz}^{y_p} c \\
\end{array} \right] \ ,
\label{G1EQ} \end{eqnarray}
where $c\equiv \cos (qa/2)$, $s \equiv \sin (qa/2)$,
and $J_{\alpha \beta}$ is the symmetric part of the
exchange tensor, and ${\bf D}$ is the Dzialoshinskii-Moriya
vector.  Note that this result explicitly requires two irreps, also
in accord with the macroscopic symmetry analysis.  It should
also be noted that due to the low site symmetry, the spin-phonon
coupling involves the derivatives of almost all elements of
the generalized exchange tensor. The next step (which is
ongoing) is to
evaluate the necessary gradients of the exchange tensor from
a first principle calculation and to incorporate the
mode structure and energy of the optical modes at zero
wavevector.    

\section{SUMMARY}
In this paper I have reviewed the symmetry analysis of the
interaction responsible of ferroelectricity in Ni$_3$V$_2$O$_8$
induced by incommensurate magnetism\cite{FERRO} and have also summarized
some results of the corresponding microscopic theory.\cite{REV,MICRO}
The symmetry has also recently been applied to some of the phases of
TbMO$_3$.\cite{TMO2}  It seems likely that these approaches will
prove useful for a number of other mutliferroics.

\vspace{0.2 in} \noindent
{\bf ACKNOWLEDGEMENTS}  I have obviously greatly profitted
from my collaborators, especially A. Aharony, C. Broholm, M. Kenzelmann,
G. Lawes, O. Entin-Wohlman, and T. Yildirim.
This work was partially supported by the
US-Israel Binational Science Foundation.


\begin{thebibliography}{10}
\bibitem{FOOT} 
Including phase factors in Fig. \ref{G4} is at our convenience.  If
the system wanted the $y$-components of the spin moments to be
in phase with the $x$ and $z$ components which were real-valued,
then $m_{sy}^{(4)}(q)$ would be pure imaginary.  The result of Eq. 
(\ref{REAL}) indicates fact that the $m_{s \alpha}$'s
of a given representation all have the same phase. But because
some of the spine spin components are proportional to $im_{s\alpha}$,
those spine spin components with factors of $i$ are
$90^{\rm o}$ out of phase from those which do not have such factors.
\end{thebibliography}
\end{document}